\documentclass[twocolumn]{revtex4}
\usepackage{graphicx}
\usepackage{dcolumn}
\usepackage{bm}
\usepackage{amsmath}
\usepackage{amsfonts}
\usepackage{amssymb}
\usepackage{color}
\usepackage{epstopdf}
\usepackage{float}
\providecommand{\U}[1]{\protect\rule{.1in}{.1in}}
\newcommand{\ket}[1]{|#1\rangle}

\newcommand{\modu}[1]{\vert\vec{#1}\vert}

\begin{document}
\title{Out-of-equilibrium quantum thermodynamics in the Bloch sphere: 
	temperature and internal entropy production}
\author{Andr\'es Vallejo}
\author{Alejandro Romanelli}
\author{Ra\'ul Donangelo}
\affiliation{\begin{small} Facultad de Ingenier\'{\i}a, Universidad 
de la Rep\'ublica, Montevideo, Uruguay\end{small}}
\date{\today}
\begin{abstract}
\noindent An explicit expression for the temperature of an open 
two-level quantum system is obtained as a function of local properties,
under the hypothesis of weak interaction with the environment. 
This temperature is defined for both equilibrium and out-of-equilibrium 
states, and coincides with the environment temperature if the system 
reaches thermal equilibrium with a heat reservoir. 
Additionally, we show that within this theoretical framework the total 
entropy production can be partitioned into two contributions: one due 
to heat transfer, and another, associated to internal irreversibilities,
related to the loss of internal coherence by the qubit.
The positiveness of the heat capacity is established, as well as its 
consistency with the well known results at thermal equilibrium. 
We apply these concepts to two different systems, and show that they 
behave in analogous ways as their classical counterparts.

\end{abstract}

%\pacs{03.67-a, 05.45Mt}
\maketitle

\section{\label{sec:level1}Introduction}

Despite the enormous success of classical thermodynamics during the first 
half century after its creation, some important thermodynamic properties 
only acquired their deepest physical interpretation when considering the 
microscopic degrees of freedom, that is to say, with the advent of 
statistical mechanics. 
Entropy, for example, is a typical case of the above: the famous Boltzmann 
equation $S=k_{B}\ln\Omega$ associates the classical entropy concept 
with the logarithm of the number of microscopic configurations that are 
compatible with the actual macroscopic state.

Something similar has occurred with the notion of temperature. 
Its intrinsic phenomenological character associated to our ability 
of distinguishing various degrees of cold and hot objects, made it 
difficult to express temperature in terms of other well-established 
physical quantities. 
The first relevant step came via the kinetic theory, which established 
the proportionality between temperature and the average kinetic energy 
of the microscopic components of the system, in a simple physical model. 
Once the relation between entropy and microscopic world was clear, the 
equation 

\begin{equation}\label{tempSM}
\frac{1}{T}=\frac{\partial S}{\partial E}
\end{equation} 
deduced in classical thermodynamics can be adopted as a mechanical
definition of temperature, for cases for which we are able to write 
the number of microscopic configurations as a function of the energy. 
For a recent review of the temperature concept in statistical mechanics,
see \cite{Puglisi}.

The statistical interpretation is important to shed light onto the 
thermodynamic theory based on classical mechanics. 
When considering quantum mechanics as the underlying theory, the 
probabilistic nature of the theory and the fact that open systems 
generally find themselves in mixed states, make the statistical approach 
not only convenient, but mandatory. 
In this case, the statistical description of the possible results of the 
measurements performed on the system is made through the reduced density 
matrix $\rho_{_{S}}$, which allows to obtain the expected values of the 
local observables by means of the rule:
\begin{equation}
\langle A\rangle=tr[\rho_{_{S}}A].
\end{equation} 
This inherently statistical behavior suggests that in the attempt of 
extending thermodynamics to the quantum regime, the natural candidates to 
occupy the role of thermodynamic properties are the expected values of 
certain operators, or, in the general case, functions of them. 
Typical examples are the internal energy, usually defined as the expected 
value of the local Hamiltonian $H_{_{S}}$, in the weak coupling regime 
(i.e., if the interaction energy can be neglected):
\begin{equation}\label{def internal energy}
E=\langle H_{_{S}}\rangle =tr[H_{_{S}}\rho_{_{S}}],
\end{equation}
and the von Neumann entropy, which can be defined as the expected value 
of the entropy operator $-\ln\rho_{_{S}}$
\begin{equation}
S_{vN}=-tr[\rho_{_{S}}\ln_{\rho_{_{S}}}].
\end{equation}
Adopting this point of view, in this work we explore the possibility of 
introducing the concept of temperature for a qubit that undergoes an open 
dynamic. 
The question about the need, convenience or even the mere possibility of 
defining the temperature of such a two-level system is not new \cite{Abragam}. 
More recently, several ``temperatures" have been proposed and demonstrated 
to be useful in order to explain certain quantum thermodynamic processes 
\cite{Romanelli1,Gemmer,Hartmann,Brunner2,Latune,Ali}. 
Unlike these approaches, some of them very sophisticated, our approach 
here is very simple and consists in explicitly applying Eq.(\ref{tempSM}), 
but employing the von Neumann entropy $S_{vN}$, instead of the 
thermodynamic entropy $S$, an idea that to the best of our knowledge, has 
not been implemented. 
Unlike $S$, which is only defined for equilibrium states, $S_{vN}$ can be 
assigned to arbitrary states. 
As a consequence, the expression obtained for the temperature is valid 
even far from equilibrium, generalizing the usual equilibrium 
temperature. 
A completely different perspective about the temperature problem in 
quantum mechanics is presented in \cite{Ghonge}, where the authors 
propose that the temperature must be considered as an operator instead of 
a realistic local variable. 
In the same direction, several temperature-energy uncertainty relations 
have been found in the context of statistical mechanics, and more 
recently in quantum mechanics \cite{Miller}.  

This work is organized as follows. In Section II, we establish a 
thermodynamic theory for the qubit based on considering the expected
values of the spin operators as the elementary thermodynamic properties. 
In Section III, we deduce the expression for the temperature of the system, 
the specific heat and the entropy production and we discuss the results. 
An analysis of the behavior of the defined quantities for some simple 
two-level models is developed in section IV. 
Finally, some remarks and conclusions are presented in Section V.

\section{Basic theory for the qubit}

\subsection{A state postulate}
In addition to the well-known four laws of thermodynamics, the classical 
theory makes use of the so-called ``state postulate", which establishes 
that the thermodynamic state of a system in equilibrium is determined by 
knowing the values of a reduced set of independent intensive properties. 
Typically, for a simple compressible system without magnetic or other 
additional effects, two independent properties completely determine the 
state; i.e. any other property is a function of those two properties \cite{Cengel}.\

In the framework of open quantum systems and following the same 
philosophy, it would be useful to choose a set of thermodynamic 
properties (in the sense discussed in the introduction) such that they 
allow to express any other one. 
Since the reduced state of an open two-level system can be put in 
correspondence with a point in the Bloch sphere, it is clear that three 
parameters are necessary to fully describe the state. 
A natural choice are the components of the Bloch vector: 
\begin{equation}
\vec{B}=(B_{x},B_{y},B_{z}).
\end{equation} 
\noindent which can be obtained as the expected values of the spin 
operators $S_{x}, S_{y}$ and $S_{z}$ (aside from a factor $\hbar/2$):
\begin{equation}
\begin{split}
&B_{x}=\langle S_{x}\rangle = tr(\rho_{_{S}}\sigma_{x})\\
&B_{y}=\langle S_{y}\rangle = tr(\rho_{_{S}}\sigma_{y})\\
&B_{z}=\langle S_{z}\rangle = tr(\rho_{_{S}}\sigma_{z}).
\end{split}
\end{equation}
\noindent where $\sigma_{x}$, $\sigma_{y}$ and $\sigma_{z}$ are the 
Pauli matrices. 
Since they are expected values of local operators, they satisfy our 
definition of a thermodynamic property. 
Additionally, they completely determine the thermodynamic state, 
considering that the reduced density matrix can be expressed in terms 
of the Bloch vector components in the following way:
\begin{equation}\label{rdm}
\rho_{_{S}}=\frac{1}{2}[1+\vec{B}.\vec{\sigma}]=\frac{1}{2}
\begin{pmatrix}
{1+B_{z}}&{B_{x}-iB_{y}}\\
{B_{x}+iB_{y}}&{1-B_{z}}
\end{pmatrix} 
\end{equation}
\noindent where $\vec{\sigma}$ is the vector whose components are the 
Pauli matrices. 
This implies that any other conceivable thermodynamic property (in 
particular the temperature) must be a function of them. 
For example, the dimensionless von Neumann entropy can be expressed in 
terms of the modulus of the Bloch vector $B$ as:
\begin{small}
\begin{equation}\label{entropy1}
\frac{S_{vN}}{k_{B}}=
-\left(\frac{1+B}{2}\right)\ln\left(\frac{1+B}{2}\right)
-\left(\frac{1-B}{2}\right)\ln\left(\frac{1-B}{2}\right) .
\end{equation}
\end{small}

\subsection{Internal energy, heat and work}
The unitary evolution of the total system (qubit plus environment) is 
governed by a Hamiltonian $H$, which in the general case can be written 
as:
\begin{equation}
H=H_{S}+H_{E}+H_{int}
\end{equation}
where $H_{S}$ is the qubit Hamiltonian, $H_{E}$ is the Hamiltonian of the 
environment, and $H_{int}$ describes the interaction between both systems.
Note that the qubit Hamiltonian $H_{S}$ can be written as a linear combination of the Pauli 
matrices, aside from a scalar multiple of the identity, without 
physical consequences since it represents a shift in the energy 
eigenvalues:
\begin{equation}\label{Hamiltonian}
H_{S}=-\vec{v}.\vec{\sigma} \;,
\end{equation}
where $\vec{v}$ is a vector which can be associated to an 
effective magnetic field.
Note also that $\pm\modu{v}$ are the eigenenergies of the system.

In the weak coupling limit, the internal energy can be 
obtained using  
Eqs.(\ref{def internal energy}), (\ref{rdm}), (\ref{Hamiltonian}), 
and with the help of the identity:
\begin{equation}\label{identity}
(\vec{a}.\vec{\sigma})(\vec{b}.\vec{\sigma})
=(\vec{a}.\vec{b})I+i\vec{\sigma}.(\vec{a}\times\vec{b}).
\end{equation}
resulting in:
\begin{equation}\label{internal energy}
E=-\vec{B}.\vec{v}
\end{equation}
i.e. the projection of the Bloch vector on the effective magnetic field.
If the Hamiltonian is time-dependent, we can write an infinitesimal 
change in the internal energy as:
\begin{equation}\label{dE}
dE=-d\vec{B}.\vec{v}-\vec{B}.d\vec{v}
\end{equation} 
As usual, we identify the energy change due to variations in the 
reduced state as \textit{heat} \textbf{\cite{Kieu}}:
\begin{equation}\label{heat}
\delta Q=-d\vec{B}.\vec{v}
\end{equation}
Since an isolated qubit undergoes an unitary evolution, 
and therefore $d\vec{B}\perp\vec{v}$, we verify that in this case 
$\delta Q=0$, as expected. 
We also identify the energy change associated to our possible 
control  over the temporal evolution of the Hamiltonian, as 
\textit{work}:
\begin{equation}\label{work}
\delta W=-\vec{B}.d\vec{v}
\end{equation}
Clearly, Eq.(\ref{dE}) can be considered as a statement of the first 
law of thermodynamics in the quantum regime:
\begin{equation}\label{first_law}
dE=\delta Q+\delta W.
\end{equation}

\section{Results}
\subsection{Temperature}

If we assume that the von Neumann entropy, Eq.(\ref{entropy1}), is a 
valid extension of the statistical entropy in the quantum regime, 
we can use Eq.(\ref{tempSM}) to obtain the temperature. 
We note that the internal energy, Eq.(\ref{internal energy}), 
can be expressed in terms of the projection 
$B_{\parallel}=\vec{B}.\hat{v}$ of the Bloch vector on 
the direction $\hat{v}$:
\begin{equation}
E=-\modu{v}B_{\parallel},
\end{equation}
and, therefore
\begin{equation}\label{deftemp}
\frac{1}{T}=- \frac{1}{\modu{v}}\frac{\partial S_{vN}}
{\partial B_{\parallel}}\bigg\rvert_{B_{\perp}}
\end{equation} 
Since $\frac{d S_{vN}}{dB}=-k_{B}\tanh^{-1}(B)$ and 
$\frac{\partial B}{\partial B_{\parallel}}\big\rvert_{B_{\perp}}=\frac{B_{\parallel}}{B}$,
the expression for the temperature is:
\begin{equation}\label{temperature1}
T=\frac{\varepsilon B}{k_{B}B_{\parallel}\tanh^{-1}(B)}
\end{equation}
where we have denoted the eigenenergy $\modu{v}$ by $\varepsilon$.
Since entropy and energy are defined for equilibrium and out of 
equilibrium states, Eq.(\ref{temperature1}) applies in both cases, 
allowing us to obtain the qubit's temperature as a function of the 
reduced state and the instantaneous Hamiltonian. 
\begin{figure}[!h]
 %\centering
{\includegraphics[width=1.0\columnwidth]{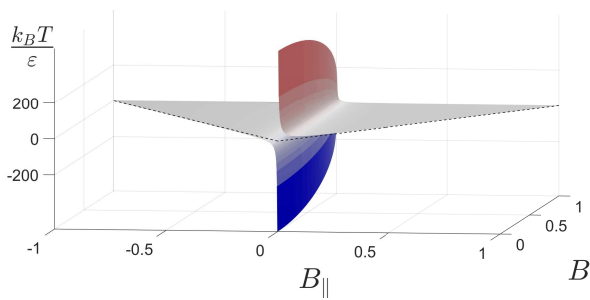}}
\caption{Dimensionless qubit's temperature $k_{B}T/\varepsilon$, 
as a function of the modulus $B$ and the $z$-component $B_{\parallel}$ 
of the Bloch vector.}
\label{fig1}
\end{figure}
 
Let us consider, as usual in physical implementations, of the qubit 
as a spin in a magnetic field pointing along the $z$ direction.
Therefore, the Hamiltonian adopts the form 
\begin{equation}\label{HS}
H_{S}=-\varepsilon\sigma_{z}. 
\end{equation}
In Fig. \ref{fig1} we plot the qubit's temperature as a function of 
$B$ and $B_{\parallel}$. 
Two branches are appreciated, one associated to positive 
temperature states (for $B_{\parallel}>0$), and the other to negative 
temperature states (for $B_{\parallel}<0$). 
This means that the spin's temperature is positive if the projection 
of the Bloch vector on the $z$ direction is aligned with the field, 
negative in the opposite case, and undefined if the expected 
value is zero (in this case, the temperature diverges to $\pm\infty$ 
as $B_{\parallel}$ goes to $0^{\pm}$).

\subsection{Consistency with the equilibrium temperature}
If the system reaches thermal equilibrium with a reservoir at positive 
temperature $T_{E}$, the time-averaged reduced state adopts the 
form \cite{Goldstein}, \cite{Popescu}:
\begin{equation}\label{thermalstate}
\overline{\rho}_{S}=\frac{e^{-\beta_{E} H_{S}}}
{tr(e^{-\beta_{E} H_{S}})},
\end{equation}
\noindent where $\beta_{E}=(k_{B}T_{E})^{-1}$. 
In this case, Eq.(\ref{rdm}) reduces to
\begin{equation}
\overline{\rho}_{S}=\frac{1}{2}[1+\vec{B_{eq}}.\vec{\sigma}],
\end{equation}
so, the Bloch vector points along the $\hat{v}$ direction and,
after some calculation we see that it has a modulus:
\begin{equation}\label{Beq}
B_{eq}=\tanh{\left(\frac{\varepsilon}{k_{B}T_{E}}\right)},
\end{equation}
so the environment temperature can be expressed in terms of the modulus 
of the equilibrium Bloch vector as:
\begin{equation}\label{temperature_env}
T_{E}=\frac{\varepsilon}{k_{B}\tanh^{-1}(B_{eq})}.
\end{equation} 
Observe that since in this case $B=B_{\parallel}=B_{eq}$, the qubit's 
temperature, Eq.(\ref{temperature1}), coincides with the environment 
temperature, Eq.(\ref{temperature_env}). 
In this case, considering Eq.(\ref{HS}), one notes that 
the natural populations, i.e., the eigenvalues of the reduced density 
matrix ($\lambda_{+/-}=1/2\pm B/2$) are the equilibrium populations 
$P_{g}$ and $P_{e}$ of the ground and excited state, respectively, 
so we re-obtain the typical relation between the environment 
temperature and the populations:
\begin{equation}\label{temp_pop}
T_{E}=T=\frac{2\varepsilon}{k_{B}\ln\left(\frac{P_{g}}{P_{e}}\right)}.
\end{equation} 
The second equality of Eq.(\ref{temp_pop}) and other similar relations are 
usually employed outside the equilibrium situation, when the introduction 
of an ``effective temperature" is useful in order to characterize the 
evolution of the system \cite{Quan,Serban,Huang,Du,Jin,Doyeux,Das,Scully,Quan1}.

\begin{figure}[!h]
 %\centering
 {\includegraphics[trim= 300 50 -50 0, scale=0.45, clip]{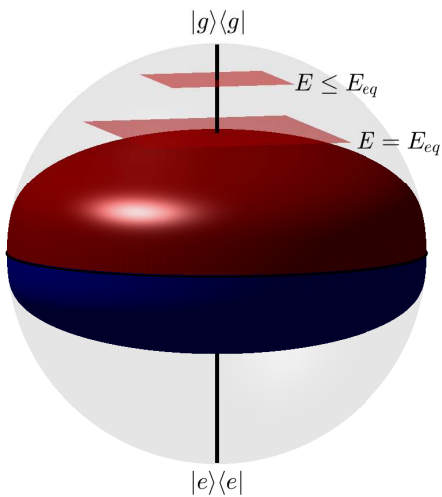}}
 \caption{Isothermal surfaces in the Bloch sphere, corresponding to the 
 values of dimensionless temperature $k_{B}T_{1}/\varepsilon$=1.5 
 	(red, northern hemisphere) and $k_{B}T_{2}/\varepsilon$ =-2 
 	(blue, southern hemisphere). Two constant energy planes are also
    shown, one tangent to the isotherm $k_{B}T_{1}/\varepsilon$=1.5, 
    which is associated to the equilibrium energy $E_{eq}$, and another
    associated to an energy smaller than $E_{eq}$ }
 \label{fig2}
\end{figure}

Fixing the temperature, Eq.(\ref{temperature_env}) allows to construct 
the isothermal surfaces on the Bloch sphere, two of which are represented 
in Fig. \ref{fig2}. 
The larger the value of the temperature, the lower the curvature of the 
corresponding isothermal surface, and as $T$ increases, the surfaces 
converge to the maximum circle located in the horizontal plane.
The surface of the sphere is also an isothermal surface, corresponding 
to temperature 0, except for the equator, located in the $z = 0$ plane, 
in which the temperature is not defined. 
There are also represented two constant energy planes. 
The lower one corresponds to the energy of the system in its thermal 
equilibrium state, $E_{eq}$, when the system is embedded in an environment 
at temperature $T_{1}$. 
Such state is located at the intersection of the corresponding isothermal 
surface with the vertical diameter. 
Note that the upper plane does not intersect the isothermal surface,
so energy values lower than that of the equilibrium at a certain temperature 
are only compatible with also lower values of the temperature.

\subsection{Internal entropy production}
At the macroscopic level irreversible processes are accompanied 
by a positive entropy production $\delta S_{gen}$, which for a  
system that undergoes an infinitesimal process exchanging only energy with the environment is defined, in terms of 
the entropy change $dS$, and the entropy flux associated to the heat 
exchanged, $\delta Q/T$, as follows:
\begin{equation}\label{second_law}
 \delta S_{gen}=dS-\frac{\delta Q}{T}
\end{equation} 
\noindent where $T$ is the temperature of the  
surface region where heat exchange takes place. 
Usually, heat reservoirs are modeled as internally reversible, which means 
that no entropy is produced in their interior. 
This implies that if we consider a control volume that includes the system 
we are studying, and whose boundary is at the temperature 
of the heat reservoir, application of Eq.(\ref{second_law}) 
allows us to obtain the total entropy generated.
This value includes the entropy produced due to irreversible heat 
transfer (if the temperature difference across the boundary is 
not infinitesimal), and eventually contributions associated to 
internal irreversibilities inside the system of interest. 
In what follows, we will see that our notion of temperature allows 
us to evaluate these contributions to the total entropy production.
  
For an arbitrary infinitesimal change in the modulus of the Bloch vector, 
from Eq.(\ref{entropy1}) the total entropy change of the qubit is:  
\begin{equation}\label{dSvN}
dS_{vN}=-k_{B}\tanh^{-1}(B)dB .
\end{equation}
Let us consider the unit vectors $\hat{B}=\vec{B}/B$ and $\hat{v}=\vec{B}/\varepsilon$. Since $dB=\hat{B}.d\vec{B}$, 
and expressing $\hat{B}$ in terms of 
its projections along the direction of $\hat{v}$ and its 
orthogonal complement, we obtain
\begin{equation}\label{dS2}
dS_{vN}=-k_{B}\tanh^{-1}(B)[(\hat{v}.\hat{B})\hat{v}+
(\hat{B}-(\hat{v}.\hat{B})\hat{v})].d\vec{B}.
\end{equation}
%Let us consider the first term on the r.h.s of the equation above. 
Using Eqs.(\ref{heat}, \ref{temperature1}), the first term on the 
r.h.s of the equation above is:
\begin{equation}
-k_{B}\tanh^{-1}(B)(\hat{v}.\hat{B})\hat{v}.d\vec{B}=\frac{\delta Q}{T}.
\end{equation}
Thus we arrive at:
\begin{equation}\label{dSgen3}
\delta S_{gen}^{int}=dS_{vN}-\frac{\delta Q}{T},
\end{equation}
where we have defined the internal entropy production $\delta S_{gen}^{int}$ as
\begin{equation}\label{sgen_int}
 \delta S_{gen}^{int}\equiv -k_{B}\tanh^{-1}(B)\left[\hat{B}-(\hat{v}.\hat{B})\hat{v}\right].d\vec{B}.
\end{equation}
Unlike the previous discussion, in this case the temperature appearing 
in Eq.(\ref{dSgen3}) is the qubit's temperature instead of the 
environment temperature, and consequently the corresponding entropy
production is only a fraction of the total entropy production, 
since the latter also includes the irreversible 
heat transfer contribution associated to the possible finite 
temperature difference between the qubit and the environment.
To clarify this point, we recall that at the quantum level the 
total entropy produced during a process is linked to the distance,
measured in a particular way, between the initial and the equilibrium 
states of the system, in case the latter exists 
\cite{Breuer, Brunelli, Vallejo2}. 
Specifically, ref.\cite{Breuer} proposes that:
\begin{equation}\label{sgenrelent}
S_{gen}^{tot}(t)=k_{B} \left[ D(\rho_{S}(0)\parallel\rho_{S}^{eq})
-D(\rho_{S}(t)\parallel\rho_{S}^{eq})\right],
\end{equation}
\noindent where 
$D(\rho\parallel\rho ')=tr(\rho\ln\rho)-tr(\rho\ln\rho ')$ 
is the Kullbak-Leibler divergence (relative entropy) of the states 
$\rho$ and $\rho '$, and $\rho_{S}^{eq}$ is the equilibrium state. 
Assuming that the equilibrium state is the thermal state at the 
environment temperature $T_{E}$, it is straightforward to see that:
\begin{equation}\label{dsgentot2}
\delta S_{gen}^{tot}=dS_{vN}-\frac{\delta Q}{T_{E}}
\end{equation}
Then, using Eqs. (\ref{dSgen3}, \ref{dsgentot2}) we obtain:
\begin{equation}\label{dsgentot3}
\delta S_{gen}^{tot}=\delta S_{gen}^{ht}+\delta S_{gen}^{int}
\end{equation}
\noindent where we have defined 
\begin{equation}\label{sgenht}
\delta S_{gen}^{ht}=\delta Q\left(\frac{1}{T}-\frac{1}{T_{E}}\right).
\end{equation}
\noindent Eq. (\ref{sgenht}) can be interpreted as the entropy production 
due the heat transfer at the system's boundary, since 
this term cancels when the temperatures of the system 
and the environment coincide, or when the heat exchanged is zero. 
This confirms that the second term of Eq. (\ref{dsgentot3}), 
$\delta S_{gen}^{int}$, must be associated exclusively to internal 
irreversibilities. 
\noindent The classification of the total entropy production in the internal and boundary contributions is the standard procedure in classical thermodynamics, and it is plausible also in quantum mechanics provided that the temperatures are well-defined. However, the existence of an intrinsic entropy generation even for such a
simple quantum system is a rather remarkable fact, and some of its 
relevant aspects are discussed in Appendix A.

\subsection{Heat capacity}

All qubit states located on a isothermal surface $T$=const. are out 
of equilibrium  except for the thermal state, for which the Bloch 
vector points in the direction of the applied field (the vertical diameter). 
On the other hand, Eq.(\ref{internal energy}) shows that the constant 
energy surfaces are horizontal planes such that energy decreases with 
height. 
Observing Fig. \ref{fig2}, it is easy to conclude that from all the
states at a given temperature, the equilibrium state is the one of 
least energy. 

In order to formalize these ideas, we proceed to obtain a general 
expression for the heat capacity of the system, as a function of 
the state. 
Typically, the heat capacity is defined considering a process in which 
the relevant work is zero (for example, a constant volume process for 
a compressible substance; or a constant magnetization process for a 
magnetic substance). 
If the direction of $\vec{v}$ is fixed, the zero work condition 
implies that the eigenenergy $\varepsilon$ is fixed, so in general 
it makes sense to define: 
\begin{equation}
C_{\varepsilon}=\left(\frac{\partial E}{\partial T}\right)_{\varepsilon}
\end{equation}
	
But the energy depends only on the $B_{\parallel}$ component, 
$E=-\varepsilon B_{\parallel}$, so we obtain:
\begin{equation}
	C_{\varepsilon}=-\varepsilon\left(\frac{\partial T}{\partial B_{\parallel}}\right)^{-1}
\end{equation}

After some algebra, the general expression for the qubit's 
heat capacity is:
\begin{equation}\label{Cv}
C_{\varepsilon}=\frac{k_{B}B(1-B^{2})
	[\tanh^{-1}(B)]^{2}B_{\parallel}^{2}}
	{\tanh^{-1}(B)(B^2-B_{\parallel}^2)(1-B^2)+BB_{\parallel}^2}
\end{equation}

It is interesting to test the behavior of a well known particular case. 
For the equilibrium states of the system at positive temperature 
we have that $B=B_{\parallel}$; in this case:
\begin{equation}
C_{\varepsilon}=k_{B}(1-B^{2})[\tanh^{-1}(B)]^2
\end{equation}

But as we have seen, the modulus of the equilibrium Bloch vector is 
defined by the environment temperature, which coincides in this case 
with the temperature of the qubit:
\begin{equation}
	B=\tanh{\left(\frac{\varepsilon}{k_{B}T}\right)}
\end{equation}
so we obtain:
\begin{equation}\label{Cveq}
	C_{\varepsilon}=k_{B}\left[\frac{\varepsilon/k_{B}T}
	{\cosh\left(\varepsilon/k_{B}T\right)}\right]^2
\end{equation}

Eq.(\ref{Cveq}) is the well known expression for the specific heat 
of a two-level system at thermal equilibrium, which shows once again 
the consistency of the theory with the equilibrium case \cite{Kim}.

\begin{figure}[!h]
	\centering
	{\includegraphics[width=1.0\columnwidth]{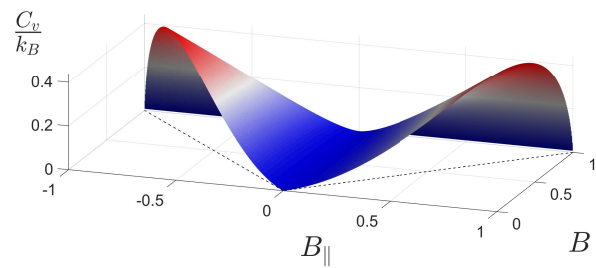}}
	\caption{Dimensionless heat capacity $C_{\varepsilon}/k_{B}$, 
	as a function of the state defined through the modulus and
	$z$ component of the Bloch vector.}\label{fig3}
\end{figure}

More in general, we note that since $\vert B_{\parallel}\vert\leq B\leq 1$, 
Eq.(\ref{Cv}) always leads to $C_{\varepsilon}\geq 0$ (see Fig. \ref{fig3}).

We also notice that pure states, and those located on the plane 
$B_{\parallel}=0$, have zero heat capacity. 
This implies that, for these states, very small energy variations 
are able to produce large temperature changes, which agrees 
with our previous observations on the temperature behavior.

Particularly remarkable also is the fact that, for any process starting 
from a zero temperature state ($B=1$), and evolving towards an infinite 
temperature state ($B_{\parallel}=0$), the value of the heat capacity 
increases from zero, reaches a maximum value, and then diminishes and 
tends asymptotically to zero. 
This effect, known as the \textit{Schottky anomaly} is well known in 
equilibrium Statistical Mechanics, and typically occurs in systems 
with a finite number of energy levels \cite{Mahdavifar}. 
The analysis of Fig. \ref{fig3} shows that the effect is present even 
in the out of equilibrium situation. In particular, $C_{\varepsilon}$ 
presents two global maxima in the regions $B=B_{\parallel}$ and 
$B=-B_{\parallel}$, with values:
\begin{equation}
	C_{\varepsilon}^{max}\simeq 0.4392\,,
\end{equation}
\noindent in agreement with the known result for a two level system 
in thermal equilibrium \cite{Kim}.

In the following section we will apply these results to some simple 
two-level models, with emphasis in the behavior of the temperature 
and the internal entropy production.

\section{Applications}
\subsection{Two two-level atoms exchanging photons}
Let us consider a system $S$ composed of two two-level atoms 
separated a distance $R$ and embedded in a common thermostat at 
zero temperature. If we focus on the case where only spontaneous 
emission is taken into account, the system undergoes a 
dissipative process governed by the Markovian master equation \cite{Jakobczyk}
\begin{equation}\label{MastEQ1}
\frac{\partial \rho}{\partial t}
=\frac{1}{2}\sum_{k,l=A,B}\gamma_{kl}\left(2\sigma_{-}^{k}
\rho\sigma_{+}^{l}
-\sigma_{+}^{k}\sigma_{-}^{}\rho-\rho\sigma_{+}^{k}\sigma_{-}^{k}\right), 
\end{equation}
\noindent where
\begin{equation}
\sigma_{\pm}^{A}=\sigma_{\pm}\otimes\mathbb{I}_{2},
\hspace{0.2cm}\sigma_{\pm}^{B}=\mathbb{I}_{2}\otimes\sigma_{\pm} 
\hspace{0.2cm}\sigma_{\pm}=\frac{1}{2}(\sigma_{x}+i\sigma_{y}).
\end{equation}
Above, $\gamma_{AA}=\gamma_{BB}=\gamma_{0}$ is the single atom 
spontaneous emission rate, and 
$\gamma_{AB}=\gamma_{BA}=\gamma=g(R)\gamma_{0}\leq\gamma_{0}$ 
is the photon-exchange relaxation constant, where the function 
$g(R)$ approaches 1 as $R\rightarrow 0$. 

The explicit solution of Eq.(\ref{MastEQ1}) for an arbitrary 
initial density matrix can be found in \cite{Jakobczyk}, 
where it is used to evaluate the level of transient 
entanglement produced between the atoms due to the photons exchange. 
Here we are interested in the thermodynamic aspects of the evolution, 
so we consider such exchange as a heat transfer process between the 
atoms, with losses to the environment. 
This is so because considering the atoms at different locations 
implies that $\gamma <\gamma_{0}$, and consequently, the system 
composed of both atoms loses energy and asymptotically relaxes 
towards the product of the ground states, $\ket{0}\otimes\ket{0}$, 
regardless of the initial state. 
This kind of process is adequate to study the behavior of the 
temperature definition proposed in this work.

\begin{figure}[!h]
 \centering
  {\label{f:temperaturaV1}
    \includegraphics[width=0.7\columnwidth]{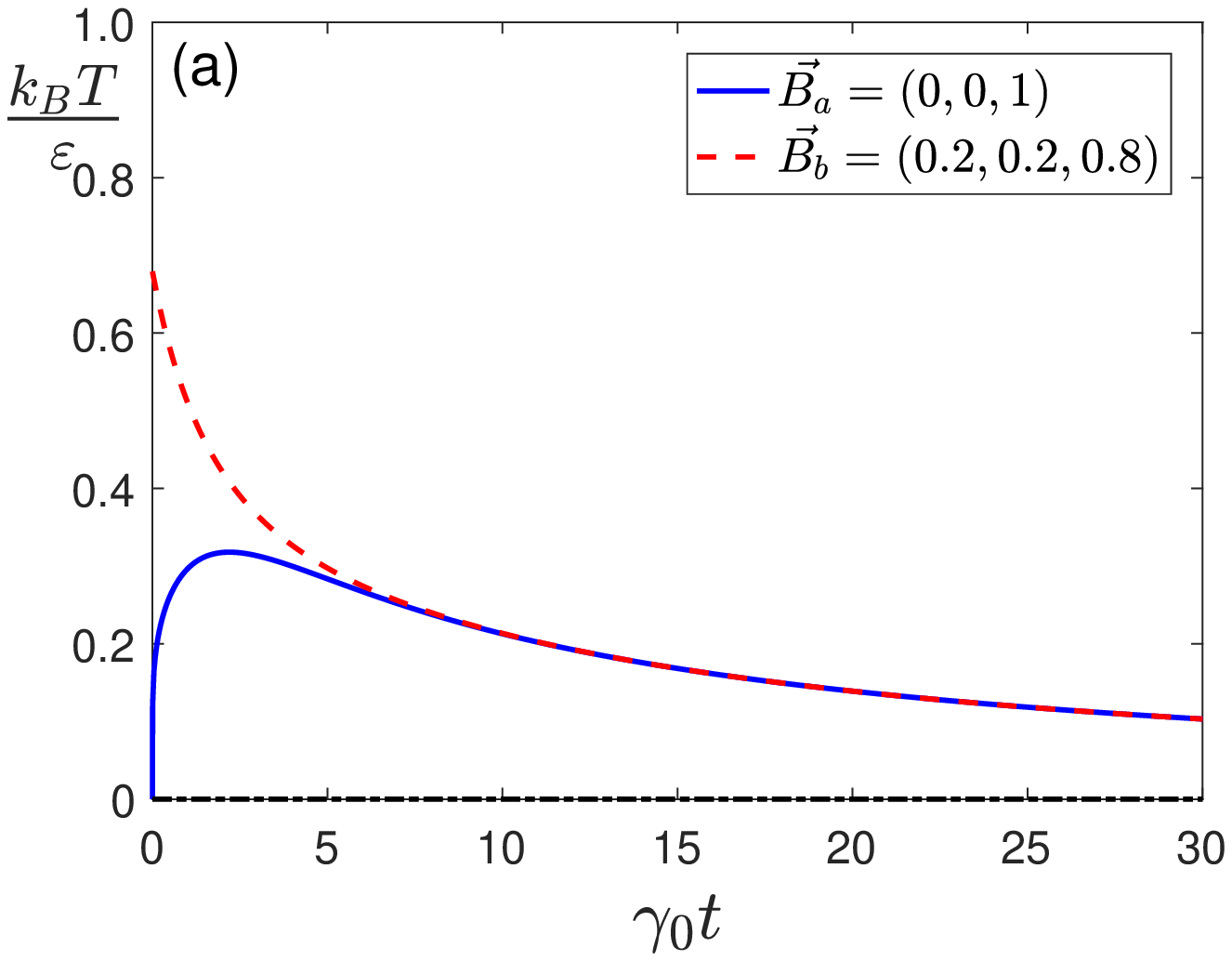}}
        %(a)
%    \subfloat[a]
  {\label{f:energiaNMV1}
    \includegraphics[width=0.7\columnwidth]{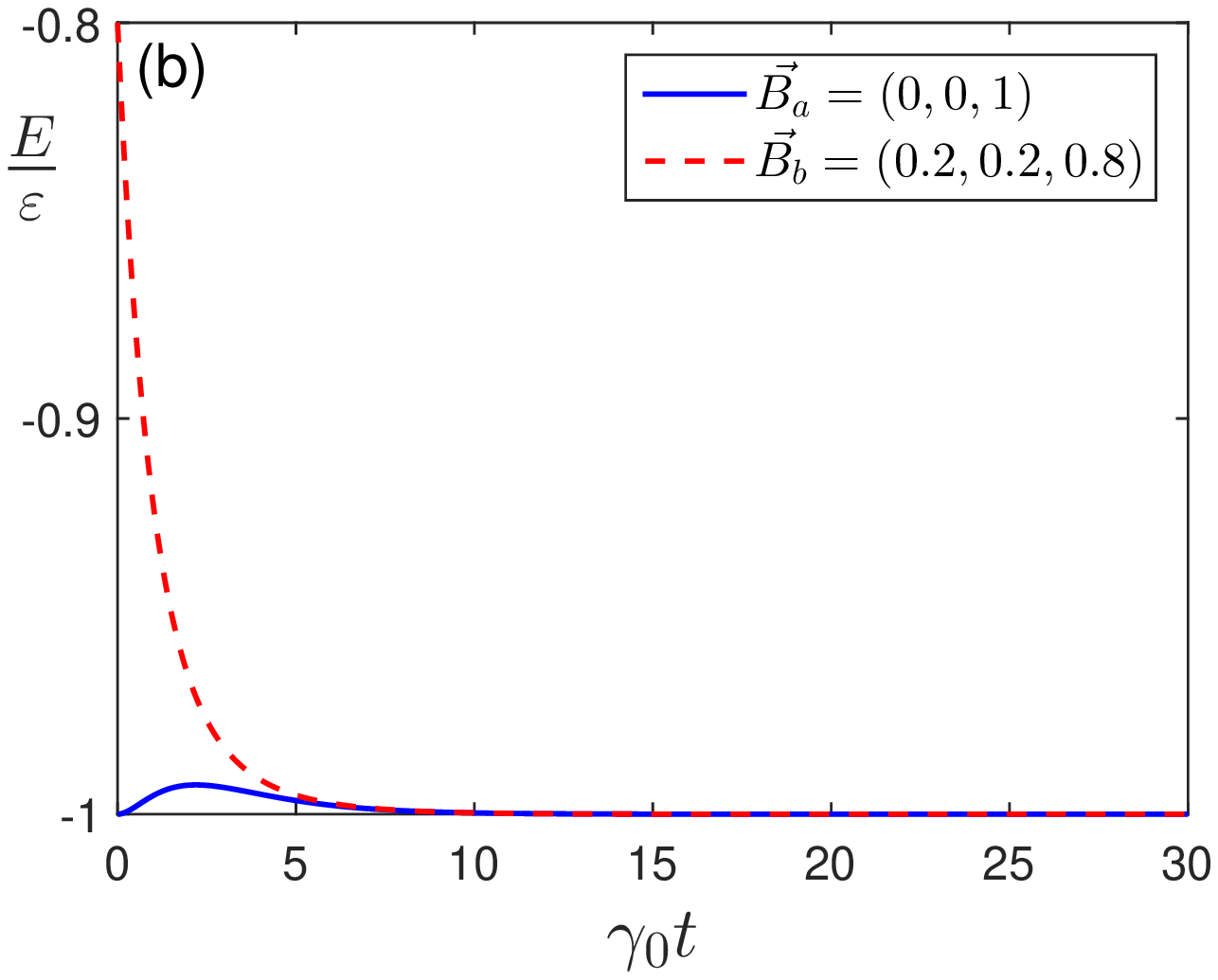}}
    %(b)
  {\label{f:entropyNMV1}
    	\includegraphics[width=0.7\columnwidth]{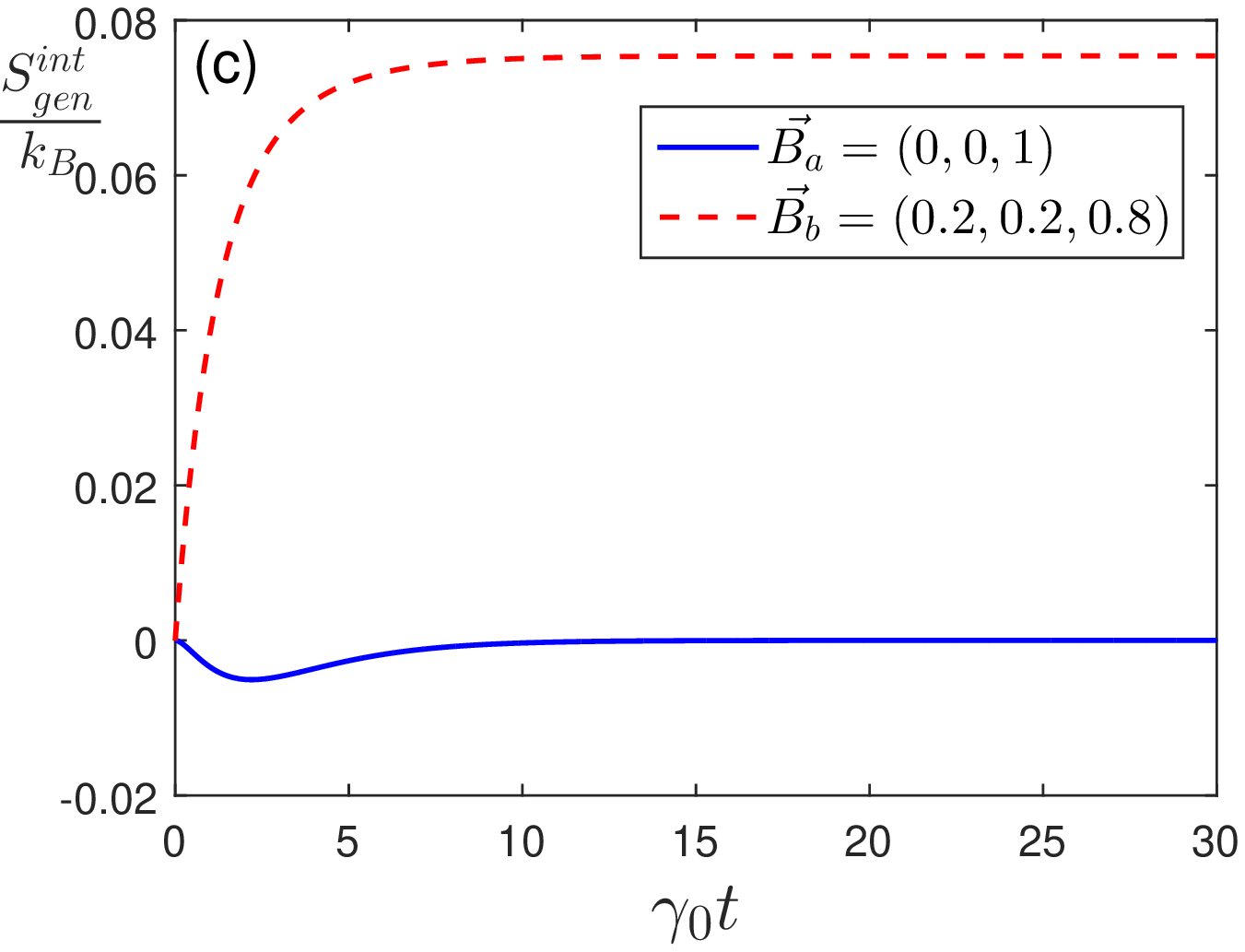}}
      
 \caption{Temperature, internal energy and entropy production time evolutions,
 	as a function of $\gamma_{0}t$ for two interacting two-level atoms
 	$a$ and $b$.\\
 	(a) Dimensionless temperature $k_{B}T/\varepsilon$.
    (b) Dimensionless internal energies $E/\varepsilon$. 
    (c) Dimensionless internal entropy production.
    The initial reduced state is a product state of local 
    densities defined by the Bloch vectors $\vec{B}_{a}=(0,0,1)$ 
    and $\vec{B}_{b}=(0.2,0.2,0.8)$, and $g(R)=\gamma/\gamma_{0}=0.5$ The environment temperature is zero (black dash-dot line.)} 
 \label{fig4}
\end{figure}

In Fig. \ref{fig4} we present the comparative time evolution of the 
temperatures and internal energies for both atoms. 
The global initial state is a product state, with atom $a$ in 
its ground state ($E_{a}/\varepsilon=-1,T_{a}=0$), and atom $b$ 
in a mixed state such that $E_{b}=-0.8,T_{2}>0$. 
Initially atom $a$ absorbs a fraction of the energy emitted by  
atom $b$, while its own emission is negligible, since it starts 
from the ground state, so both its energy and temperature grow. 
Meanwhile, atom $b$ releases energy and its temperature decreases. 
At some point, $T_{a}$ is large enough, and $T_{b}$ low enough, 
so that the emission by atom $a$ exceeds its absorption, so $T_{a}$ 
and $E_{a}$ reach a maximum value simultaneously and subsequently 
start to decrease. 
Finally, we observe that thermal equilibrium between the atoms 
occurs before thermal equilibrium with the environment. 
In fact, once $T_{a}$ equals $T_{b}$, no energy flux occurs 
between the atoms and the composed system behaves as a unique system 
since the temperature and energy of both atoms are the same for all 
subsequent times. 
From this moment on, they cool together and thermalize with 
the environment. 
This behavior is reminiscent of the one that undergo two macroscopic 
bodies at different temperatures in thermal contact between them, 
and with an environment at a temperature less than or equal to 
the lowest of those of the systems involved.

Note that, although the internal entropy generated associated to
atom $a$ has a negative initial transient, the sum of the 
contributions of both atoms is monotonically increasing.
The decrease in entropy of atom $a$ is associated to the
fact that it was initially in thermal equilibrium, so when
it interacts with atom $b$ it starts moving away from the 
$z$-axis. Since the variation of the distance to this axis 
is a measure of the internal entropy generated (see Appendix A),
this explains the excursion into negative entropy values in
the case of this atom. 

\subsection{Two-level system interacting with a heat reservoir: 
the Jaynes-Cummings model}
The interaction between a two-level atom and the electromagnetic 
field is described by the multi-mode Jaynes-Cummings Hamiltonian:
\begin{equation}
H=\frac{\hbar}{2}\omega_{0}\sigma_{z}
+\hbar\sum_{k}\left[\omega_{k}b_{k}^{\dag}b_{k}
+\lambda_{k}\sigma_{+}b_{k}+\lambda_{k}^{*}\sigma_{-}b_{k}^{\dag}\right]
\end{equation}
\noindent where $b_{k}$ and $b_{k}^{\dag}$ are the creation and 
annihilation operators associated to the \textit{k}th mode of 
the field with frequency $\omega_{k}$ and coupling constant 
$\lambda_{k}$, and $w_{0}$ is the atomic transition frequency.

\begin{figure}[!h]
\centering
  {\label{f:temperaturaNMV1}
    \includegraphics[width=0.7\columnwidth]{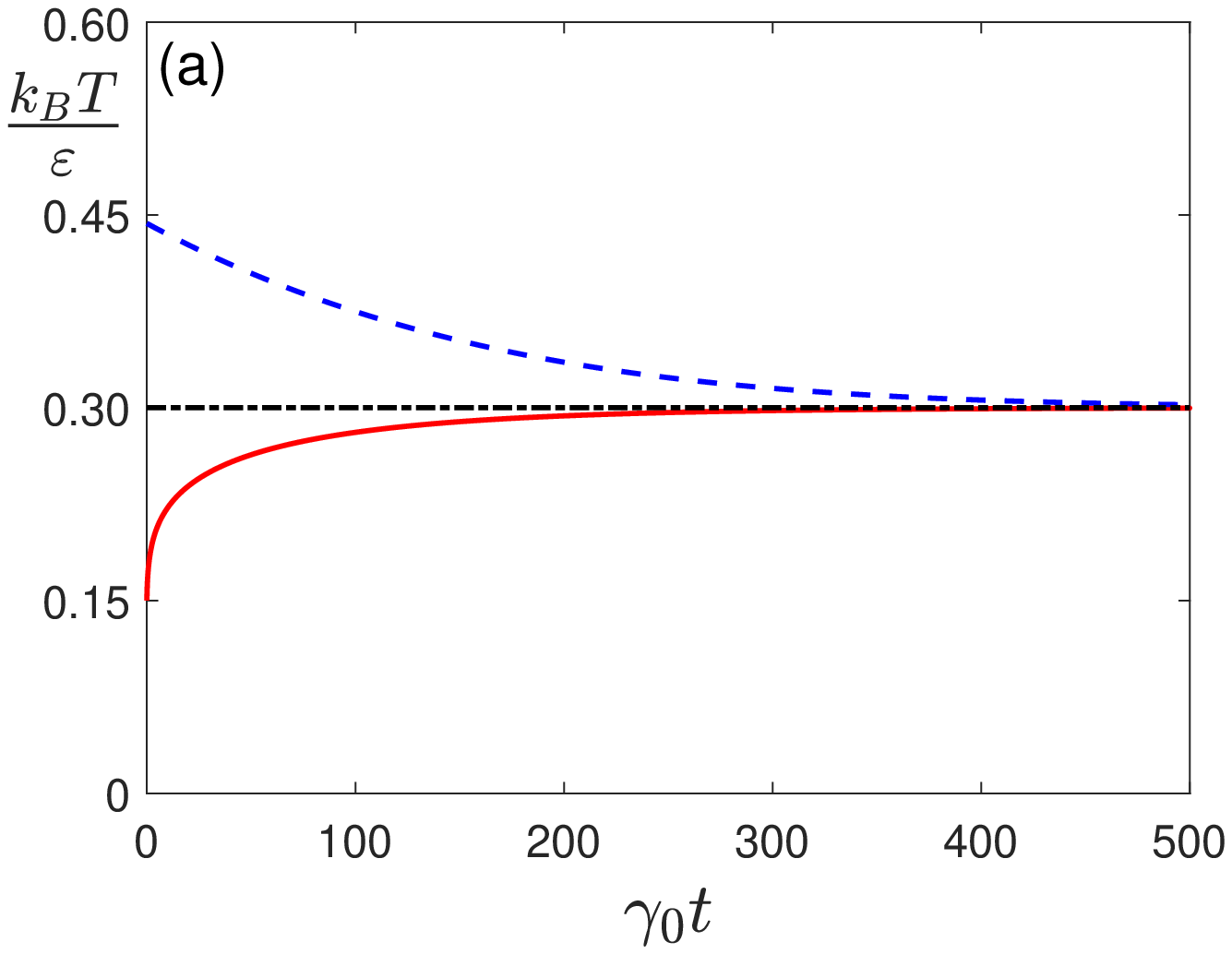}}
  {\label{f:energiaNMV1}
    \includegraphics[width=0.7\columnwidth]{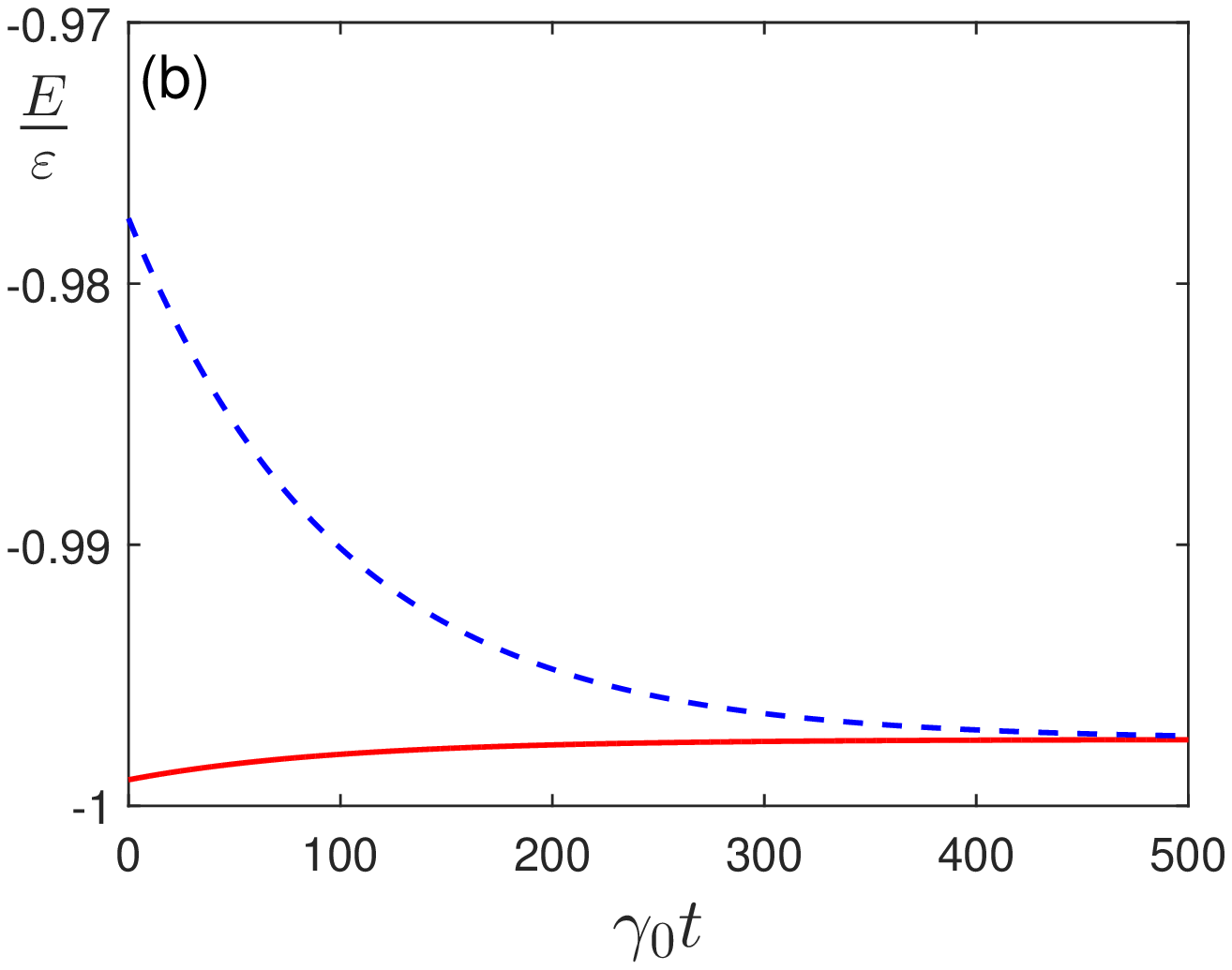}}
      {\label{f:sgenintNMV1}
    \includegraphics[width=0.7\columnwidth]{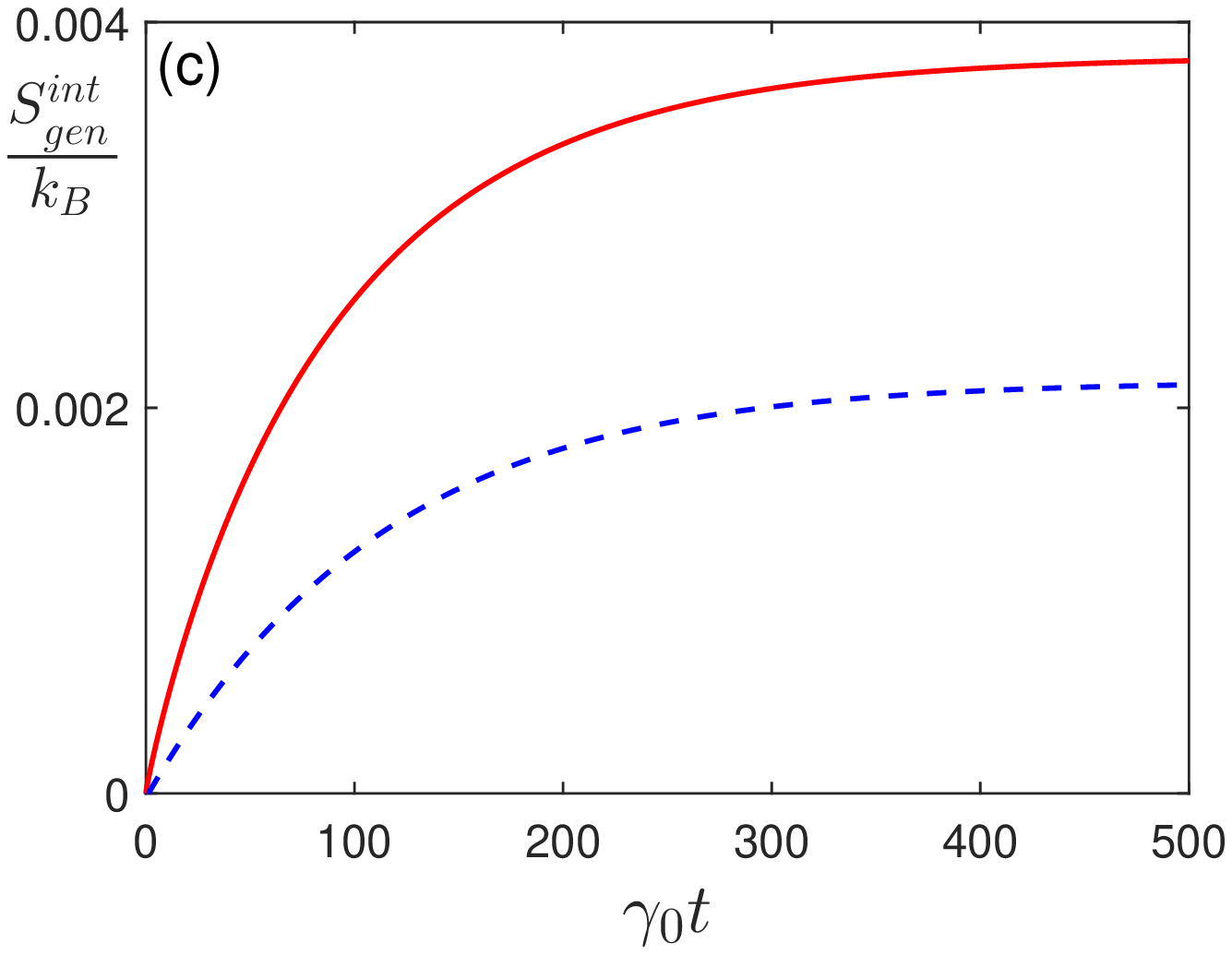}}
 \caption{Temperature, internal energy and entropy time evolutions
 	for a two level-system interacting with a heat reservoir.
 	(a) Dimensionless temperature $\frac{k_{B}T}{\varepsilon}$, 
 	as a function of $\gamma_{0}t$.
    (b) Dimensionless internal energy $\frac{E}{\varepsilon}$.
    (c) Internal entropy production, in units of $k_{B}$.
     The temperatures of the initial states considered are 
     0.5$T_{eq}$ (red full line) and $1.5T_{eq}$ (blue dashed line). 
     The initial environment temperature is $k_{B}T_{e}/\varepsilon=0.15$. The equilibrium temperature of the system is represented by the black dash-dot line}
 \label{fig5}
\end{figure}

For thermal radiation at temperature $T_{E}$, explicit expressions 
for the elements of the atom's reduced density matrix, in the 
weak coupling and low temperature approximation, can be found 
in \cite{Shresta}. 
The analysis of this expression shows that, independently of 
the initial state, the atom reaches thermal equilibrium with 
the field at the temperature
\begin{equation}
k_{B}T_{eq}=\frac{\hbar\omega_{0}}
{\tanh^{-1}(1-2\exp{(-\frac{\hbar\omega_{0}}{k_{B}T_{E}})})}.
\end{equation}
The transition to equilibrium is shown in Fig. \ref{fig5}, 
where we plot the atom's temperature, its internal energy 
and the internal entropy production. 
Since the temperature of the environment changes during the 
evolution in an unknown way, the evaluation of the total 
entropy production cannot be performed. 
Nevertheless, the growing behavior of the internal entropy 
production can be appreciated.

More interesting conclusions arise if one studies the system 
under the Markovian approximation, valid in the limit of high 
temperature. 
In reference \cite{Breuer} it is shown that the master equation 
for the atom in the interaction picture is:
\begin{align}
{\frac{\partial\rho_{_{S}}}{\partial t}}&=
\gamma_0(\mathcal{N}+1)
\left(\sigma_{_-}\rho\sigma_{_+}-\frac{1}{2}\sigma_{_+}\sigma_{_-}
\rho-\frac{1}{2}\rho\sigma_{_+}\sigma_{_-}\right)\,\notag \\
+&\gamma_0\mathcal{N}\left(\sigma_{_+}\rho\sigma_{_-}-
\frac{1}{2}\sigma_{_-}\sigma_{_+}\rho-
\frac{1}{2}\rho\sigma_{_-}\sigma_{_+}\right) ,\, \label{scho2}
\end{align}
\noindent where $\gamma_{0}$ is the spontaneous emission rate, 
and  $\mathcal{N}$ is the Planck distribution at the transition 
frequency $\omega_{0}$:
\begin{equation}
\mathcal{N}=\frac{1}{e^{\beta_{E}\hbar\omega_{0}}-1}
\end{equation}   

The master equation can be solved with the help of Pauli matrices 
algebra, and the explicit components of the Bloch vector can be 
found in \cite{Breuer}.
With the time dependence of the three basic thermodynamic properties 
at hand, we can implement the thermodynamic analysis of the model.
\begin{figure}[!h]
 \centering
  {\label{f:SvN}
    \includegraphics[width=0.75\columnwidth]{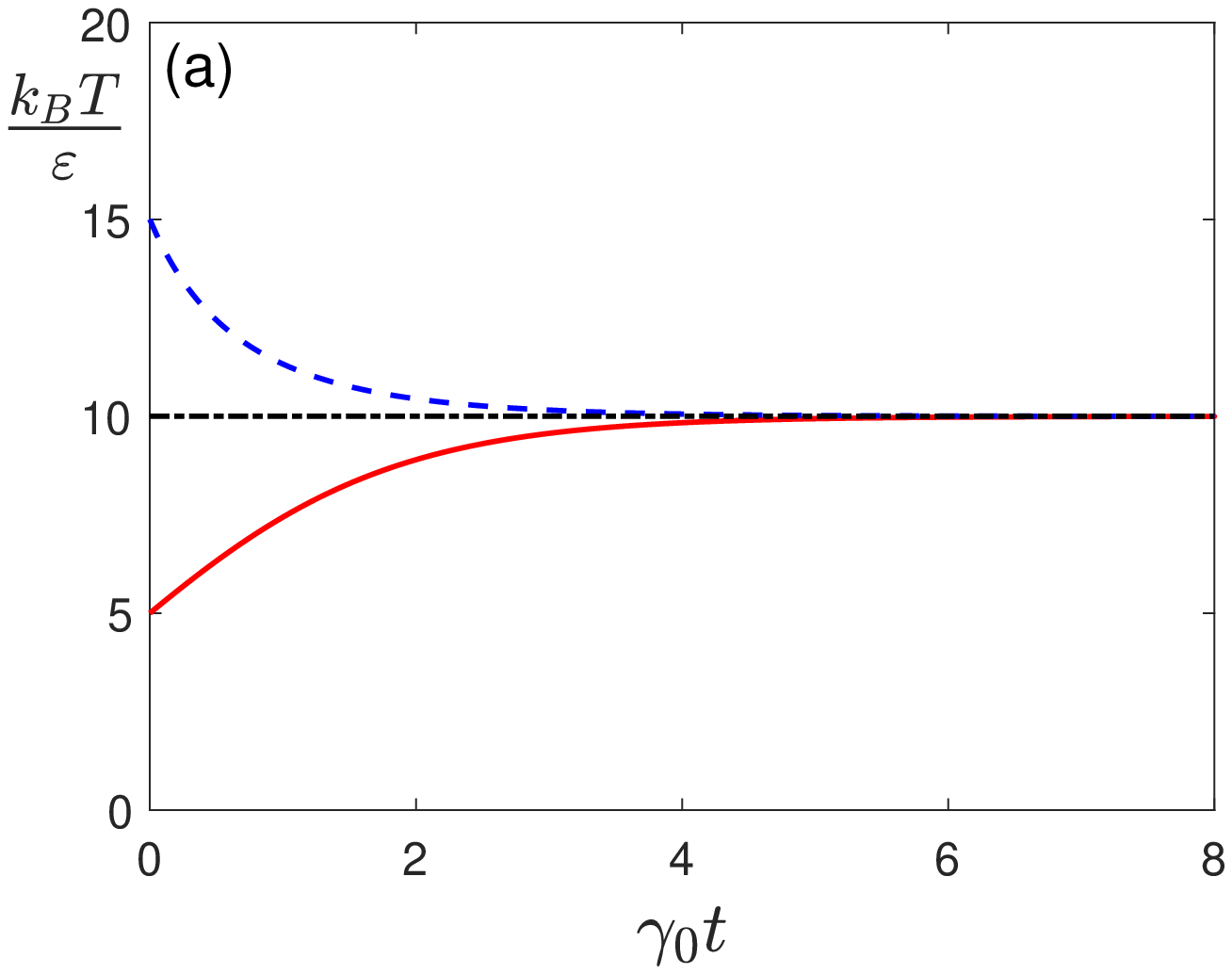}}
  {\label{f:heat_overT}
    \includegraphics[width=0.75\columnwidth]{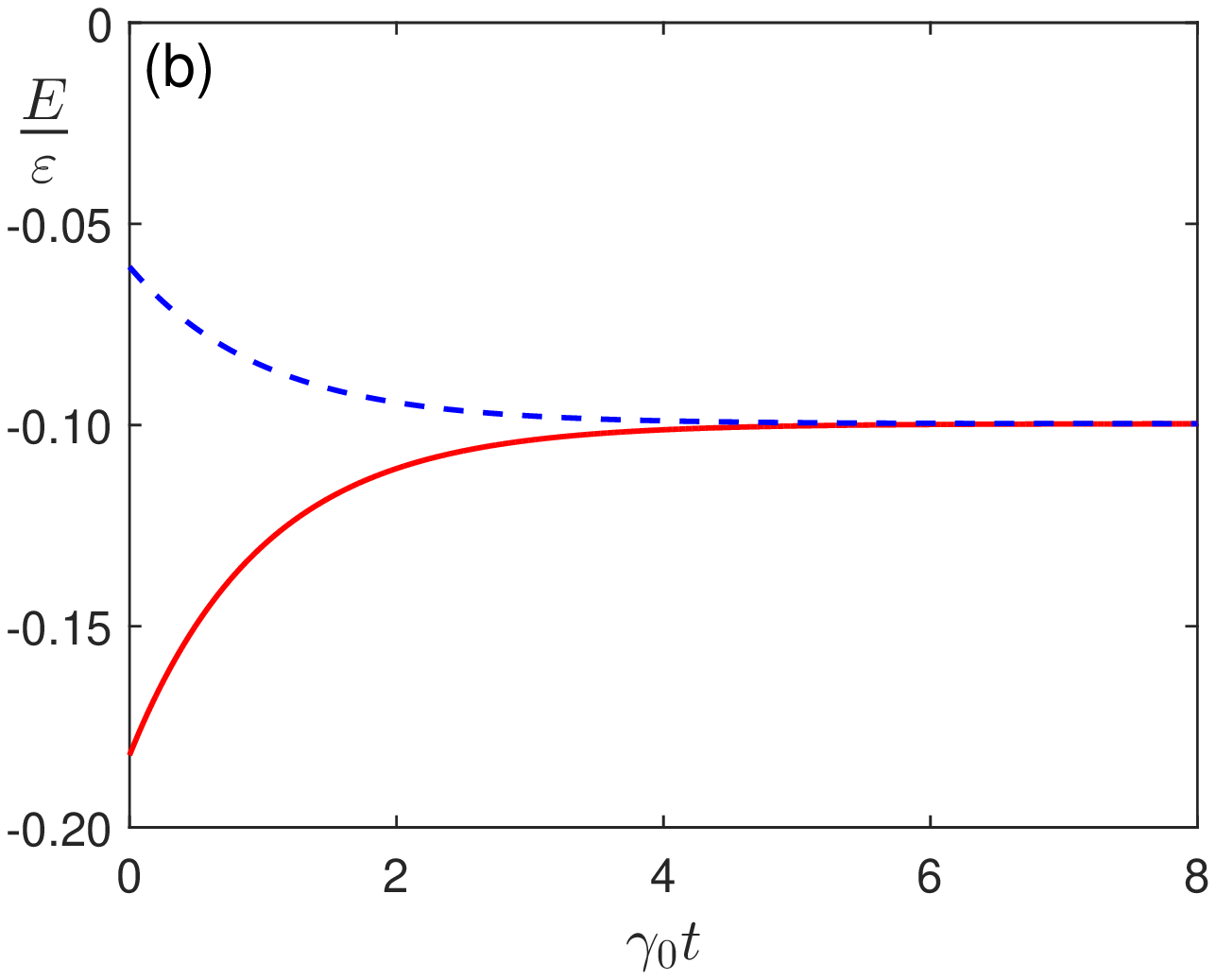}}
 \caption{Temperature and internal energy evolution of a two-level
 	system interacting with the reservoir within the Markovian 
 	approximation.
 	(a) Dimensionless temperature $\frac{k_{B}T}{\varepsilon}$.
    (b) Dimensionless internal energy $\frac{E}{\varepsilon}$, as a 
    function of $\gamma_{0}t$. 
    The time evolution of the internal entropy generated for 
    this system is shown in Fig. \ref{fig6}a.
    The dimensionless environment temperature is 10 (black dash-dot line), and the initial 
    states considered are selected in such a way that the qubit's 
    initial dimensionless temperatures are 5 (red full line) 
    and 15 (blue dashed line). 
    }
 \label{fig6}
\end{figure}

In Fig. \ref{fig6} we plot the dimensionless atom's temperature, 
Eq.(\ref{temperature1}), and the dimensionless internal energy, 
for two different initial states. 
Note that if the initial temperature is greater than the environment 
temperature $T_{E}$, the internal energy decreases in time. 
This implies that the system releases heat to the environment. 
On the other hand, for an initial temperature lower that $T_{E}$, 
the internal energy increases, so the system receives heat from 
the reservoir. 
As expected, when the asymptotic state is reached, in all cases 
the qubit's temperature coincides with the environment temperature. 
These observations leads to consider Eq.(\ref{temperature1}) as a 
real indicator of how ``warm" the qubit finds itself in the 
corresponding reduced state.

\begin{figure}[H]
	\centering
	{\label{f:SvN}
		\includegraphics[width=0.75\columnwidth]{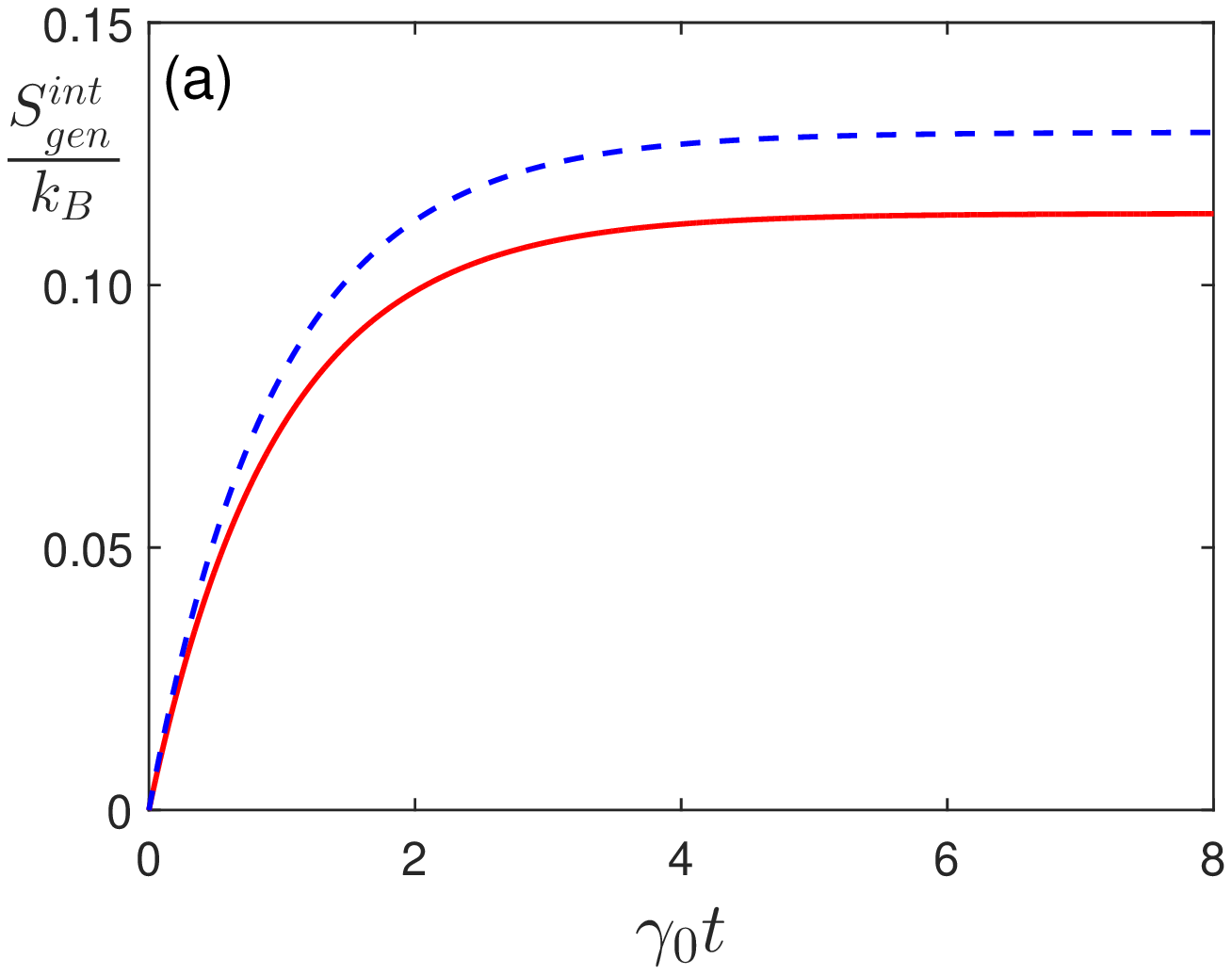}}
	{\label{f:heat_overT}
		\includegraphics[width=0.75\columnwidth]{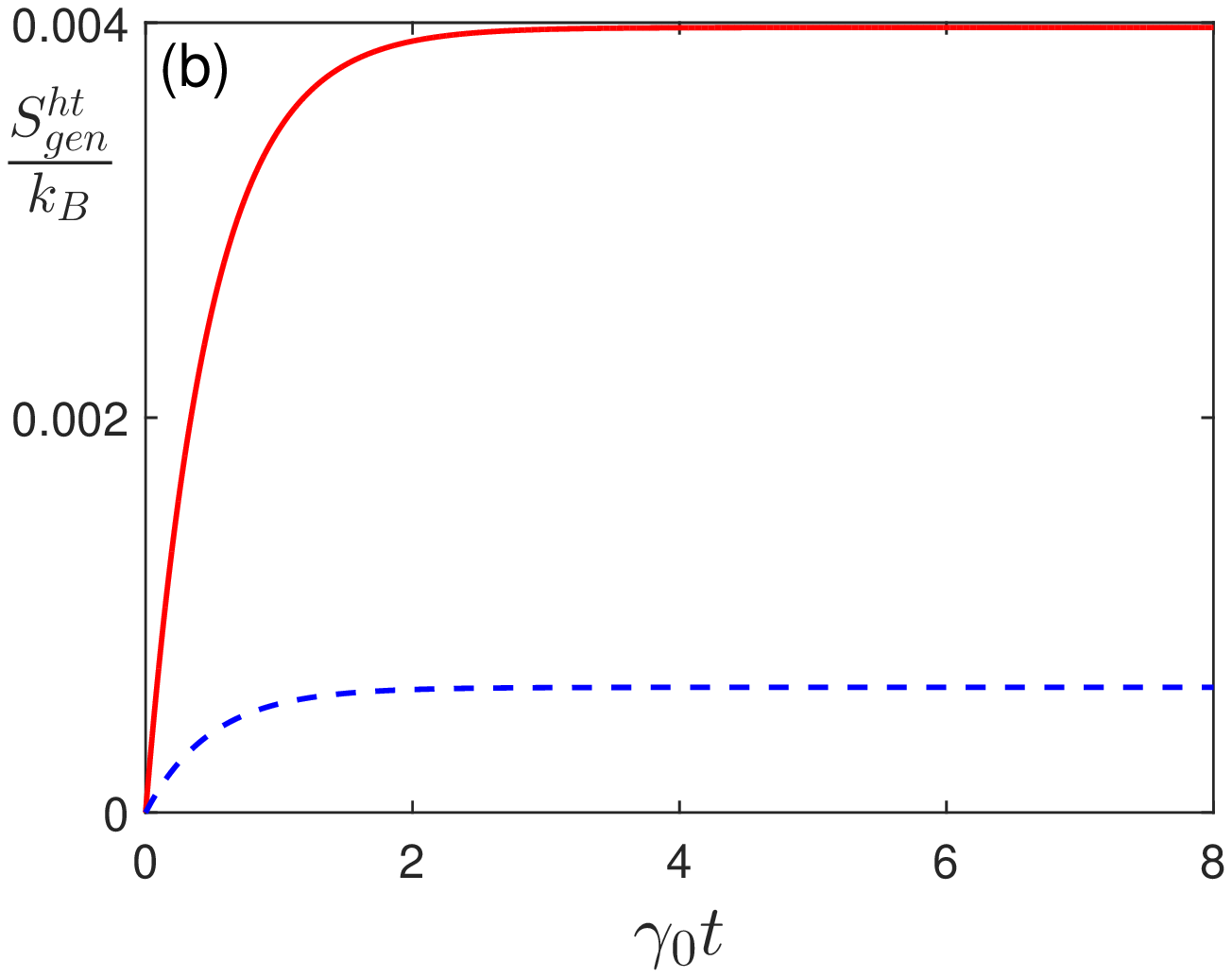}}
	{\label{f:sgen}
		\includegraphics[width=0.75\columnwidth]{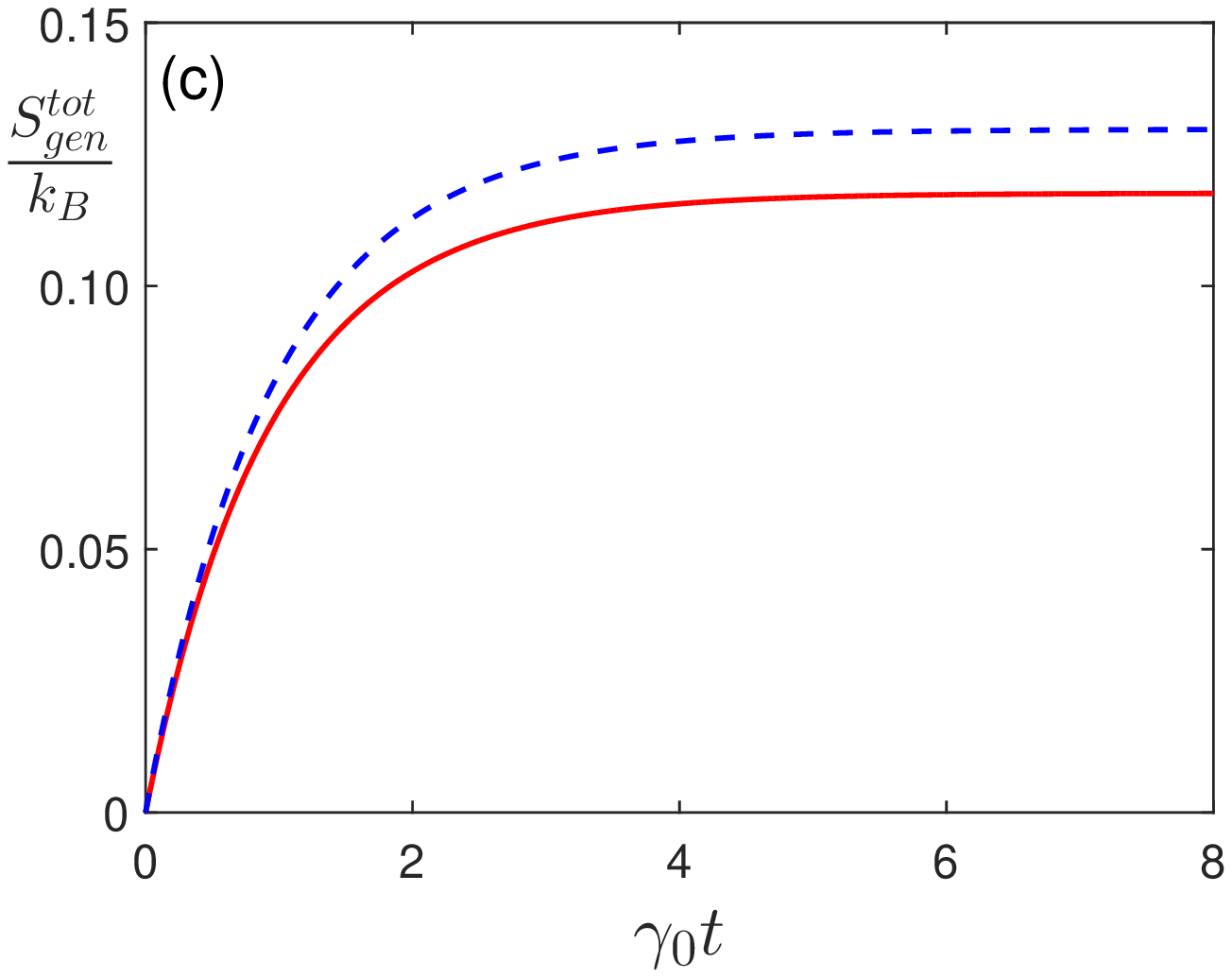}}
	\caption{Components of the entropy production for a two-level 
		system interacting with a heat reservoir.
		(a) Internal entropy production.
		(b) Entropy production due to heat transfer.
		(c) Total entropy production, in units of $k_{B}$.  
		The initial states and respective illustration codes 
		are the same of Fig. \ref{fig6}}
	\label{fig7}
\end{figure}

In this case, a complete entropy balance is possible since the 
environment remains at constant, finite  temperature during the process. 
This allows to separate the total entropy production into its 
internal (Eq.(\ref{sgen_int}), Fig. \ref{fig7}a) and boundary 
(Eq.(\ref{sgenht}), Fig. \ref{fig7}b)) contributions, as well
as their sum (Fig. \ref{fig7}c). 
The monotonically increasing behavior of these three quantities is 
verified in Fig. \ref{fig7}.   

\begin{figure}[!h]
  {\includegraphics[trim= 300 50 -50 0, scale=0.45, clip]{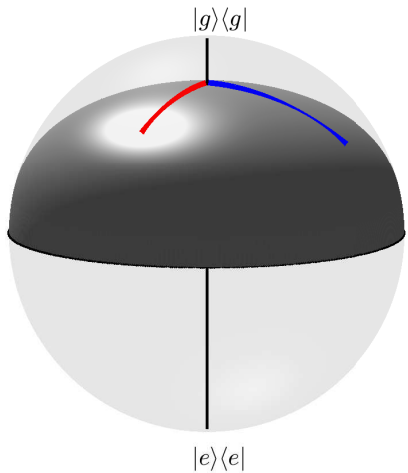}}
 \caption{Isothermal trajectories in the Bloch sphere. 
 The trajectories represented correspond to the initial value 
 $k_{B}T/\varepsilon=1$, and remain very close to the isothermal 
 surface. 
 For large $t$ they reach the thermal state located on the $z$ axis.}
 \label{fig8}
\end{figure}

The situation in which the system starts with the same temperature 
as the environment but in an out-of-equilibrium state was also studied. 
For all initial states considered satisfying this condition, 
the evolution of the system is essentially isothermal, as can be 
seen in Fig. \ref{fig8}.  
This results, as in the classical case, in the reversibility of 
the heat transfer, and as a consequence, that the boundary 
contribution to the total entropy production is negligible. 
In this case, the growing behavior of the total entropy 
production, and of the correlations between the system and the bath, 
can be understood in a purely geometrically way, since they are due 
exclusively to the tendency of the Bloch vector to point along the 
z direction, reducing the Euclidean distance between the point 
representing the reduced state, and the z axis.
 
The fact that systems in out of equilibrium states with a temperature 
given by Eq.(\ref{temperature1}) that coincides with the environment 
temperature, evolve on isothermal surfaces is rather remarkable, 
and in our opinion is a strong argument in favor of adopting that 
expression as the qubit's temperature.

\section{Final Remarks and conclusions}
In this work we have presented a natural extension of the temperature 
concept, valid for two-dimensional open quantum systems in generic
out-of-equilibrium processes, under the hypothesis of weak interaction 
with the environment. 
The construction is based on the fact that the von Neumann entropy and 
the expected value of the Hamiltonian are well-defined in the 
out-of-equilibrium situation. 

The use of the von Neumann entropy guarantees consistency with the 
equilibrium temperature when thermal equilibrium states are considered. 
In the model systems analyzed, this temperature has a behavior that is 
reminiscent of the classical behavior, correctly indicating the senses 
of the heat fluxes between initially uncorrelated systems.

We have observed that the adoption of this temperature
allows to identify the internal and heat transfer contributions to 
the total entropy production. 
Although the geometrical interpretation of the internal entropy 
generated is clear, the understanding of the physical reasons 
associated to the existence of internal irreversibilities requires 
further analysis, as well as its interpretation from the 
information-theoretical perspective.
A first step in this study is given in Appendix A1, where we
show that, for a system of two qubits, the total entropy 
generated can be directly put into correspondence with the 
change in quantum mutual information between the qubits. 
Furthermore, in Appendix A2 we show that the internal entropy 
contribution to the total generated entropy can be measured 
through the loss of quantum coherence in each of the qubits.

We have also obtained an explicit expression for 
the heat capacity of the system as a function of the state. 
The resulting expression is non-negative for all possible states, 
assuring the standard relation between heat fluxes and temperature 
when no work is involved. 
It also allows to recover the results obtained in the thermal 
equilibrium situation, such as the Schottky anomaly, and to show 
the existence of a similar behavior outside equilibrium.

The exploration of this theoretical framework for other two-level 
models is currently under investigation, as well as its implications 
in relevant tasks, such as work extraction. 
Generalizations to quantum systems with a more complex structure 
are also been considered.

\section*{Acknowledgments}
This work was partially supported by CAP, ANII and PEDECIBA (Uruguay).
\appendix
\section{About the internal entropy production, Eq. (\ref{sgen_int})}
\subsection{Entropy production and correlations in a two-qubit system}
Note that Eq.(\ref{dsgentot3}) does not include 
internal entropy production associated to the reservoir. 
It is an approximation valid in the limit of a large heat bath, 
so that it is reasonable to assume that it remains in the canonical state 
at temperature $T_{E}$ during the process (which is consistent 
with the internal reversibility hypothesis of thermal reservoirs 
in macroscopic systems). 
But in the case of interactions with a finite environment, 
the existence of an additional term of internal entropy 
production is expected. This can be illustrated considering the 
limit case: a system composed of two qubits.
We start from the following identity for the von Neumann entropy 
(denoted by $S$) of a bipartite system AB:  
\begin{equation}\label{Svn identity}
S_{A}+S_{B}-S_{AB}=I(A:B)\geq 0,
\end{equation}
\noindent where $I(A:B)$ is the \textit{quantum mutual information}:
\begin{equation}\label{mutual information}
I(A:B)=D(\rho_{AB}\parallel \rho_{A}\otimes\rho_{B}),
\end{equation}
and represents a measure of the total correlations 
(quantum and classical) between $A$ and $B$ \cite{Nielsen}. 

Each system satisfies the relation:
\begin{equation}\label{second law 2}
dS_{A/B}=\frac{\delta Q_{A/B}}{T_{A/B}}+\delta S_{gen (A/B)}^{int},
\end{equation}
\noindent so integrating in time and replacing in Eq.(\ref{Svn identity}), 
we obtain:
\begin{equation}
\begin{split}
&\int_{0}^{t}\frac{\delta Q_{A}}{T_{A}}+S_{gen(A)}^{int}(t)
+\int_{0}^{t}\frac{\delta Q_{B}}{T_{B}}+S_{gen (B)}^{int}(t)\\ &=I(A:B)(t)-[S_{A}(0)+S_{B}(0)-S_{AB}(t)].
\end{split}
\end{equation}
Since entropy is preserved by the unitary evolution of the global 
system, we have that $S_{AB}(t)=S_{AB}(0)$, and we can identify 
the terms between brackets on the r.h.s of the last equation as 
the initial mutual information: \
$I(A:B)(0)=S_{A}(0)+S_{B}(0)-S_{AB}(0)$. 
Then, using that $\delta Q_{A}=-\delta Q_{B}=\delta Q$, it follows that:
\begin{equation}
\begin{split}
S_{gen(A)}^{int}(t)+&+\int_{0}^{t}\delta Q\left(\frac{1}{{T_{A}}}-
\frac{1}{T_{B}}\right)+S_{gen (B)}^{int}(t)\\&=I(A:B)(t)-I(A:B)(0).
\end{split}
\end{equation}
As before, we can interpret the second term as the entropy 
production due to heat transfer, $S_{gen}^{ht}$. 
So we conclude that the total entropy production can be 
obtained by adding the internal contributions produced in 
each system plus the boundary term, and it equals the 
change of mutual information:
\begin{equation}\label{SgenI}
S_{gen}^{tot}=S_{gen(A)}^{int}+S_{gen(B)}^{int}+S_{gen}^{ht}=\Delta I(A:B).
\end{equation}
Eq.(\ref{SgenI}) establishes, for the present system, 
the equivalence between the production of entropy and 
the creation of correlations. 
The relation between both phenomena has been reported in Ref. \cite{Esposito}, 
for a system placed in contact with one (or several) reservoirs, each one 
of them in a thermal state.   

Observe that if $I(A:B)(0)=0$, i.e., in absence of initial correlations, 
the non-negativity of the mutual information leads to:
\begin{equation}
S_{gen}^{tot}=S_{gen(A)}^{int}+S_{gen(B)}^{int}+S_{gen}^{ht}\geq 0
\end{equation}
\noindent and the non-negativity of the entropy production is guaranteed. 
The observation that previous correlations (i.e., before the systems 
are placed in interaction) must be negligible for the second law to be 
true in its classical form can be tracked to Boltzmann himself \cite{Vitagliano}.

Nevertheless, in the case of initially correlated systems, it is known 
that mutual information can decrease, and violations of the classical 
statements of the second law are expected. 
In particular, the existence of an anomalous heat flow (from a low 
temperature to a high temperature) has been predicted \cite{Partovi,Jennings}, 
and experimentally demonstrated in the case of a two-qubit system, prepared 
in a correlated initial state such that the marginal states are thermal \cite{Micadei}. 
Such anomalous heat flow does not necessarily imply an inversion of the 
arrow of time, since the system can be understood as acting as a refrigerator, 
using the work potential stored in the correlations \cite{Bera}. 
Moreover, it has been shown that no system-bath correlations are necessary 
in order to reverse the heat flow, but only internal correlations 
associated to quantum coherences \cite{Latune2}.

The interplay between correlations, entropy production and work extraction 
in quantum systems is an active field of research, with major implications 
in our theoretical understanding of the physical world \cite{Skrzypczyk,Parrondo,Brandao,Goold,Alipour,Manzano}.

\subsection{Internal entropy production as coherence loss}
In order to give a physical interpretation to the internal entropy
production, it is convenient to express  Eq.(\ref{sgen_int}) in 
spherical coordinates. 
Choosing the z axis in the direction $\hat{v}$ (now considered fixed), 
and setting 
$d\vec{B}=dB\hat{B}+Bd\theta\hat{e_{\theta}}
+B\sin{\theta}d\varphi\hat{e_{\varphi}}$, 
we obtain:
\begin{equation}
\delta S_{gen}^{int}=-k_{B}\tanh^{-1}(B)\sin{\theta}[dB\sin{\theta}+B\cos{\theta}d\theta]
\end{equation}
Since $dB\sin{\theta}+B\cos{\theta}d\theta=d(B\sin\theta)$, 
and $B\sin\theta$ is the euclidean distance between the point 
representing the reduced state and the $z$ axis, which coincides 
with the value of the coherence of the state measured using the $l_{1}$
norm, $C_{l_{1}}$ \cite{Baumgratz}: 
\begin{equation}
C_{l_{1}}\equiv\sum_{i\neq j}\vert \rho_{S_{ij}} \vert=B\sin\theta, 
\end{equation} 
we can write:
\begin{equation}\label{sgen_int_2}
\delta S_{gen}^{int}=-k_{B}\tanh^{-1}(B)\sin{\theta}dC_{l_{1}}.
\end{equation} 
We note that in the case that the system evolves over incoherent states,
$\vec{B}\propto \vec{v}$, $\sin\theta=0$, no entropy is produced. 
In this case total reversibility is not assured, since the 
temperatures of the system and the environment may be different, 
which would imply an irreversible heat transfer. 
However, if the system is initially in thermal equilibrium, and the 
temperature of the environment changes slowly enough for the system 
to stay in thermal equilibrium throughout the process, the heat 
transfer contribution to the total entropy production will be also zero. 
Such processes are frequently employed in the theoretical construction 
of quantum  power cycles \cite{Huang2}. 

A second case of interest is when the evolution is unitary. 
In this case the Bloch vector rotates around $\hat{v}$ at constant 
angular speed, keeping its modulus and angle $\theta$ fixed. 
This implies that the coherence is constant, $dC_{l_{1}}=0$, 
and consequently, $\delta S_{gen}^{int}=0$, as in the previous case.
 
These two cases suggest that internal reversibility can be 
considered as equivalent to the preservation of the coherence. 
On the other hand, we note, considering Eq.(\ref{sgen_int_2}), 
that whenever coherence is lost, $dC_{l_{1}}<0$, entropy is produced 
$\delta S_{gen}^{int}>0$. 
This is typically the case when the system evolves towards an 
equilibrium state. 
The relation between entropy production and quantum coherence has been recently studied from a more general perspective in \cite{Santos}.

\end{document}